\documentclass[twocolumn,10pt]{asme2e}

\usepackage{amsmath}
\usepackage{xcolor}
\usepackage{graphicx}
\usepackage{caption}

\usepackage[compact]{titlesec}
\titlespacing{\section}{0pt}{1ex}{1ex}
\titlespacing{\subsection}{0pt}{1ex}{0ex}
\titlespacing{\subsubsection}{0pt}{0.5ex}{0ex}

\usepackage{hyperref}

\hypersetup{
    citebordercolor = {1 1 1},
    colorlinks,
    linkcolor={red!50!black},
    citecolor={blue!50!black},
    urlcolor={blue!80!black}
}

\confshortname{IMECE 2021}
\conffullname{the ASME 2021 International Mechanical Engineering Congress 
              Exposition}

\confdate{August 30-September 2}
\confyear{2021}
\confcity{Virtual}
\confcountry{Online}

\papernum{IMECE2021-70759}

\title{Dynamic Placement of Rapidly Deployable Mobile Sensor Robots Using Machine Learning and Expected Value of Information}

\author{Alice Agogino \thanks{Address all correspondence to this author.} \\
    Email: agogino@berkeley.edu \\
    {\tensfb Hae Young Jang} \\ 
    {\tensfb Vivek Rao}
    \affiliation{
	Dept. of Mechanical Engineering\\
	University of California\\
	Berkeley, CA 94720\\
    }	
}

\author{Ritik Batra \\
    {\tensfb Felicity Liao} \\
    {\tensfb Rohan Sood}
    \affiliation{Dept. of  Electrical Engineering \\
    and Computer Sciences\\
	University of California\\
	Berkeley, CA 94720\\
    }
}

\author{Irving Fang
    \affiliation{
    Division of Computing, \\
    Data Science, and Society \\
    Dept. of  Mathematics\\
	University of California\\
	Berkeley, CA 94720\\
    }
}

\author{R. Lily Hu
    \affiliation{
    Google Research\\
    Google\\
	Mountain View, CA 94043\\
    }
}

\author{Emerson Shoichet-Bartus
    \affiliation{Division of Computing, \\
    Data Science, and Society \\
	University of California\\
	Berkeley, CA 94720\\
}
}

\author{John Matranga
    \affiliation{Business Incubation and Acceleration\\ AVEVA\\
	San Leandro, CA 94577\\
    }
}

\begin{document}

\maketitle
\begin{abstract}
\textit{Although the Industrial Internet of Things has increased the number of sensors permanently installed in industrial plants, there will be gaps in coverage due to broken sensors or sparse density in very large plants, such as in the petrochemical industry. Modern emergency response operations are beginning to use Small Unmanned Aerial Systems (sUAS) that have the ability to drop sensor robots to precise locations. sUAS can provide longer-term persistent monitoring that aerial drones are unable to provide. Despite the relatively low cost of these assets, the choice of which robotic sensing systems to deploy to which part of an industrial process in a complex plant environment during emergency response remains challenging.} 

\textit{ This paper describes a framework for optimizing the deployment of emergency sensors as a preliminary step towards realizing the responsiveness of robots in disaster circumstances. AI techniques (Long short-term memory, 1-dimensional convolutional neural network, logistic regression, and random forest) identify regions where sensors would be most valued without requiring humans to enter the potentially dangerous area. In the case study described, the cost function for optimization considers costs of false-positive and false-negative errors. Decisions on mitigation include implementing repairs or shutting down the plant. The Expected Value of Information (EVI) is used to identify the most valuable type and location of physical sensors to be deployed to increase the decision-analytic value of a sensor network. This method is applied to a case study using the Tennessee Eastman process data set of a chemical plant, and we discuss implications of our findings for operation, distribution, and decision-making of sensors in plant emergency and resilience scenarios.}
\end{abstract}
\section*{1. INTRODUCTION}
The Industrial Internet of Things (IIoT), with enabling sensors and communication capabilities, are critical in the effective operation and management of modern industrial systems \cite{kulkarni2005knowledge, 10.1115/IMECE2017-71829, 10.1115/IMECE2018-88262}. Specifically, techniques of data collection and analysis serve as an extension of human operators and allows for improved risk management systems.

However, sensors themselves are also subject to faults or damage, and thus are key sources of plant risks. Installed sensors introduced within industrial systems are designed to be the central support for any risk detection and response strategies \cite{4796311}. Thus, when malfunctioning, these sensors fail the risk management process and endanger the industrial system entirely.

The decision making process behind planning, supporting, and deploying sensors to minimize faults and risks involves a challenging balance between cost and performance \cite{10.1115/1.2951943,10.1115/1.2830221}. Current decision-making approaches have focused on cabled or wireless systems; the former can be expensive and inflexible, while the latter can have challenges with accuracy and connectivity \cite{savazzi2013wireless}. As a result, industrial system planners often over-deploy sensors to ensure plant coverage through overcompensation. This approach presents its own challenges, specifically finding the optimal trade-off point between the cost of the deployed sensors and response, as well as the overall effectiveness of risk management \cite{7588228}. For large operations, it may not be cost effective to install sensors over large areas, particularly in remote geographical regions.

Mobile sensors accompanied by a rigorous decision-making process could alleviate some of these challenges by expanding flexibility and range \cite{wang2016}. In particular, a mobile sensor network can support a variety of industrial systems through fitting the unique optimal node placements for each process design. Fewer sensors can be deployed by identifying and covering the key features for risk detection. In addition, the placements can be continuously iterated upon should the particular industrial environment or operating process change. Thus, mobile sensors could serve as effective extensions of human operators in managing and maintaining industrial plants. To realize this use of mobile sensors, we must take the preliminary step of exploring sensor systems and decisions \cite{SquishyWeb}. 

In this paper, we propose a system for decisions on sensor placement for industrial resilience, using (1) a mobile robotic sensor platform and (2) dynamic calculation of the expected value of sample information (EVSI) from placing sensors. Dynamic placement is defined as supporting rapid, responsive, and sequential deployment of new or supplemental sensors. At each iteration, the system evaluates which sensors would produce the most valuable information, and directs mobile sensor platforms to fulfill that need. We illustrate this approach by applying it to the Tennessee-Eastman Process data set\cite{downs1993plant} \cite{TEextended}, using the results of applying machine learning methods in the calculation of EVSI. This exploration is a preliminary  step towards realizing the responsiveness of robots in disaster situations. 

This work is guided by two research questions: 

\begin{itemize}
\item R1. How can rapidly deployable mobile sensor robots be used to make data-driven decisions about which sensors to deploy in emergency situations? 
\item R2. How can machine learning approaches support the calculation of the expected value of information when faults are uncertain? 
\end{itemize}

In this paper, we first cover key related work on industrial sensing for decision support and the use of expected value of information (EVI) in resource allocation (Section 2). We then describe our methods (Section 3), summarize results (Section 4) and discuss these results (Section 5). We conclude with a discussion of future research.

\section*{2. BACKGROUND AND RELATED WORK}
\subsection*{2.1 Machine Learning and Robotics Platforms for Industrial Sensing}

Rapidly deployable mobile sensor platforms offer a promising avenue to address challenges with incomplete sensor coverage or sensor failures with permanently installed sensors \cite{wang2016, BALLARI2012102, schmidt_smith_hite_mattingly_azmy_rajan_goldhahn_2019}. Emerging mobile sensor platforms include aerial sensors in drones, sensor robots that can be deployed from aerial vehicles, or ground-based mobile sensor robots. \cite{SquishyWeb, 8968086, 6387986,7487415}. These robots can greatly enhance the performance of their sensors with the utilization of search state algorithms, for instance, dynamic programming search \cite{933092, 6387986}. Within mobile sensor platforms, droppable soft robot platforms are particularly advantageous over humans in industrial environments due to their safety and adaptability to a wide range of application areas. Previous work covering soft robotics has explored their applications in space exploration, education, and defense \cite{10.1115/1.4036014}.

 With sufficient data from industrial sensors, machine learning can be used to improve data analytics over time for enhanced robot monitoring, diagnostics, and early warning of potential failures. Machine learning can also compensate for limitations in the sensor measurements, response to environmental conditions, and changes in system parameters \cite{Luo2018TensegrityRL}. This work extends on previous studies in mobile robotic sensing platforms by proposing a workflow to integrate mobile sensing platforms with machine learning and EVI calculations in the context of the benchmark Tennessee Eastman process.


\subsection*{2.2 Fault Detection in Industrial Systems}

 Fault detection is a crucial goal of industrial sensing systems. Reliable fault detection techniques are required to ensure and maintain system stability and thus increasing operational uptime \cite{he2007fault}. Many approaches to fault detection have been proposed, including quantitative model-based, qualitative model-based and process history-based strategies \cite{venkatasubramanian2003review, park2020review}. Among these approaches, machine learning (ML) methods have emerged as an area of great interest for their efficiency and accuracy in combining sensor data with expert knowledge to predict and detect system faults. Specifically, the feature-selection aspect of ML methods can be configured to greatly reduce the need for manual configuration and minimize redundancy in its predictions \cite{hu2016machine, HU2019117}. 

For industrial applications, the Tennessee Eastman (TE) process data set has been a useful resource for testing the ability to detect and predict faults \cite{UDUGAMA202020}. In particular, the TE process can be simulated under normal operating condition as well as 21 faulty conditions. Thus, the data produced are widely used as a benchmark in industrial process control and fault detection \cite{TEextended}\cite{TEData2019}. Several efforts have been made to propose novel ML-based fault detection approaches using the TE process data set \cite{yin2014study}. These have shown differing accuracy rates in fault detection for different models and techniques.  Consequently, ML fault detection is continuously improved upon through alternative methods or techniques such as prior knowledge incorporation \cite{kulkarni2005knowledge, HU2019117}. Central to fault detection approaches leveraging ML methods is the appropriate identification and selection of features for dimensional reduction of data, widely practiced in the work examining industrial plant processes \cite{verron2008fault, chiang2004fault, 9444425} and in ML approaches more broadly \cite{7324139,cai2018feature}. 

This work extends on previous studies of ML and fault detection by coupling ML methods with Expected Value of Information calculations (see next section). We use ML to determine which sensors are most useful in the TE process data set to detect faults, using a subtractive approach among sensors during model training. This effectively adapts feature identification and dimensional reduction to a plant-specific scenario. Based on model performance, we are able to inform the calculation of expected value of information for sequentially positioning sensors in fault detection. Though basing the ML analysis on the TE process data, this procedure illustrates how industrial systems can lean on ML and mobile sensors for fault detection. With further incorporation of location-specific sensor and fault data, the effective ML methods explored can be applied to real industrial systems.


\subsection*{2.3 Expected Value of Information}
Fault detection and resolution in industrial sensing systems often rely on sparse data. In such situations, the presence of new information may aid in improved situational awareness and the resolution of potential faults. Expected value of information (EVI) is a decision-analytic approach used to ascertain when investment in discovering new information is likely useful and cost-effective, especially in data-sparse situations \cite{https://doi.org/10.1111/j.1539-6924.1999.tb00395.x}. EVI depends on the prior distribution of currently-available information, and hence, is formulated as a Bayesian approach \cite{ijcai2017-215}. While the expected value of perfect information (EVPI) is usually not achievable from a practical standpoint, the expected value of sample information (EVSI) is used to determine whether new observations lead to an increase in utility \cite{doi:10.1177/0272989X07302555}.

In recent years, the use of EVI to perform efficiency analysis has been widely used in health economics and related fields. It is frequently used to determine optimal sample sizes for randomized clinical trials based on the EVI of the results of clinical trials \cite{doi:10.1177/1740774508098413}. Another common application is in medical research, where EVSI is used to determine whether additional research should be conducted on a topic \cite{doi:10.1177/0272989X09344746}.

EVSI has also been applied to sensing. Maximizing the global EVSI of all data can be used to produce an optimal spatial placement of sensors \cite{Sensor, BALLARI2012102}. Such approaches iteratively add new observations via sequential sampling, at each step choosing the observation corresponding to the largest increase in global EVSI \cite{Sensor, doi:10.1287/ijoc.1090.0327}.

This work builds on the theoretical probabilistic models of EVSI. The successes of similar methods in medical economics and medical research lay the groundwork for a practical application of EVSI to the novel field of robotic mobile sensing \cite{Memarzadeh2016ValueOI, MALINGS2016219}. We define a posterior probability distribution and use EVSI to describe the expected benefit of sensor data collected by a rapidly deployable mobile robot in addition to existing data from a traditional industrial sensing system. Machine learning methods are then utilized to determine the gain in fault detection accuracy. This approach shares similarities with reduction in feature selection, 

Beyond this work, by moving past the simulation data set and incorporating site-specific data of industrial systems, we can achieve real world uses of EVSI and machine learning for fault detection.  
\section*{3. METHODS}
This section summarizes the opportunities afforded by new rapidly deployable mobile sensor robots and the proposed EVSI framework for sensor deployment prioritization. The data set used to illustrate the framework is described along with the associated decision tree and EVSI model for prioritizing the sensor deployment. Please refer to our Github repository (\url{https://github.com/BerkeleyExpertSystemTechnologiesLab/EVSIvsLSTM}) for the sampled dataset in CSV form, code and models.

\begin{figure}[t!]
\centering
\includegraphics[height=0.9\columnwidth]{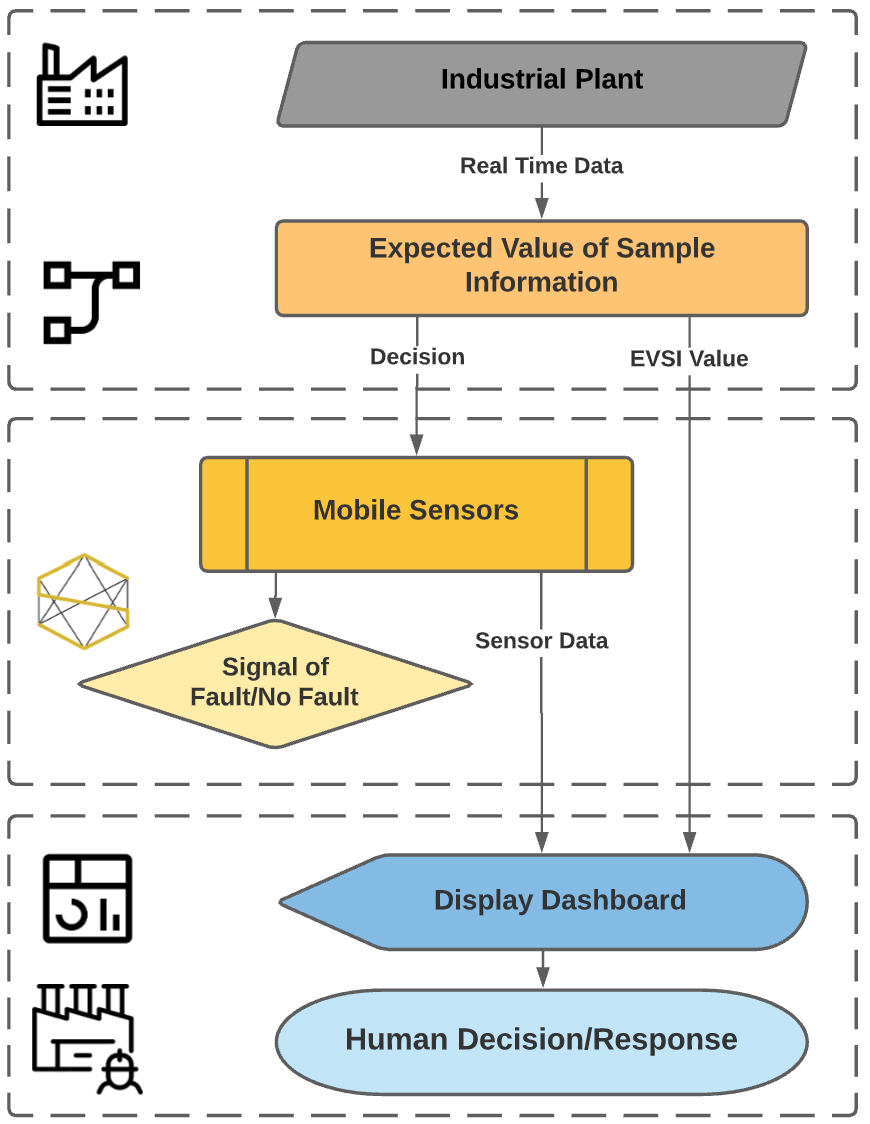}
\caption{Information flow in proposed system.}
\label{information_flow}
\end{figure}

\subsection*{3.1 Rapidly Deployable Mobile Sensor Robots} 

A spin-off of research on space exploration with NASA, the University of California at Berkeley (UC Berkeley) and Squishy Robotics have developed rapidly deployable mobile sensing robots for disaster response and remote monitoring \cite{10.1115/1.4036014}. They enable life-saving maneuvers and secure the safety of first responders by providing situational awareness and sensor data in uncharted terrain. These light-weight, low cost, robust mobile platforms can survive large impacts while carrying a payload of delicate sensors \cite{SquishyWeb}. These robots can measure and send data from dangerous environments that expose health risks to humans. They are equipped with visual, audio, chemical, biological, radiological, acoustic, GPS, and thermal sensors\cite{SquishyWeb}. Edge computing provides rapid information to first responders when and where it is needed.  The value of information collected from these robots is enhanced by use of sensor fusion and machine learning algorithms over time.  The platform is customizable and can be used for diagnostic, monitoring and prognostics in industrial situations. Although they can't replace all human functions, when combined with drones, these sensor robots can be deployed with a customized sensor payload designed for a wide range of sensor and mobility capabilities to provide improved ground-aerial situation awareness and afford appropriate placement based on EVSI \cite{Sensor}. The envisioned role of mobile sensing robots in a decision-making framework related to plant operations is shown in Fig. 1.

\subsection*{3.2 TE Process Data Set and Machine Learning Approaches} 

To detect faults, we train machine learning models on the Tennessee Eastman (TE) data set \cite{TEextended}. The TE data set corresponds to a chemical plant and the data includes a total of 52 measured and manipulated variables along with labelled faults. Measured variables ($X_{i}$) are detectable by sensors whereas manipulated variables ($M_{j}$) are controlled by an operator. In our dataset, we have 41 measured variables which are observed by sensors and 11 manipulated ones, which can be adjusted by an operator. Note, the TE data set does not contain information regarding the locations of its constituent sensors. To best model the data and predict disturbances, we considered both measured and manipulated variables as input features of interest. 

We randomly sampled from the provided training, validation, and test sets in the original TE dataset \cite{TEextended}. Please refer to Appendix B for the details on the TE dataset and our sampling methods. By framing this as a supervised machine learning problem, we approached the data set with four potential machine learning algorithms: Long short-term memory (LSTM)\cite{hochreiter1997long}, 1-dimensional convolutional neural network (CNN) \cite{fukushima1982neocognitron}, logistic regression \cite{cox1958regression}, and random forest \cite{breiman2001random}. We chose these four as they represent of a range of varying complex machine learning algorithms which enable us to observe the optimal complexity via comparisons in performance. Details of our machine learning approach are provided in Appendix A. Using all 52 manipulated and measured variables, the LSTM produces the highest accuracy and precision on the validation set and was therefore selected as the model to use in subsequent calculations of the EVSI. This comparison is further quantified in a later section.

Next, we used our model to uncover which variables most impacted the validation accuracy and would therefore be the most integral sensors to be deployed in a potential emergency situation. The validation accuracy is the accuracy of the model on the validation set which was 20\% of the TE data set. 

\begin{figure}[t!]
\centering
\includegraphics[width=1\columnwidth]{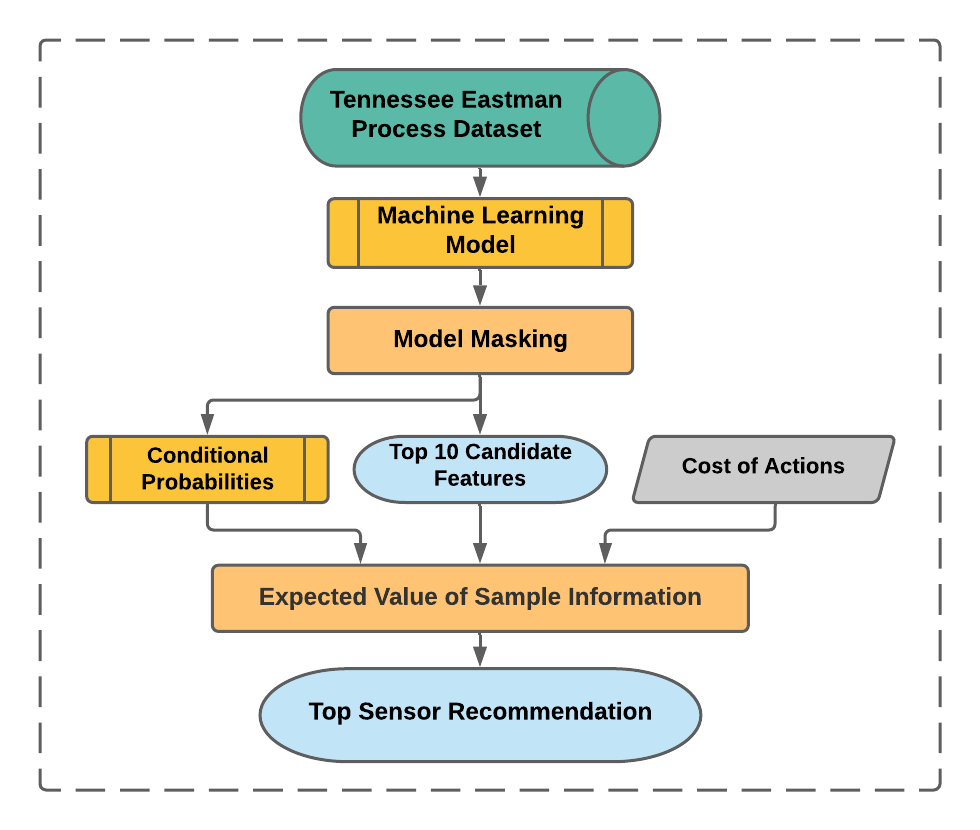}
\caption{Process for selecting top 10 most impactful features with schematic  relating  top  features,  ML  conditional probabilities,  and  action  cost  to  EVI  and  sensor  recommendation.}
\label{process_overview}
\end{figure}

Using LSTM to retrieve the 10 most impactful variables required us to iterate through each feature and measure how much better or worse the model performs with the variable removed. We choose the best variable for an iteration based on how much the accuracy dropped on the validation set as a result of removing a variable. We call this process the \textit{masking procedure} since we are masking out a variable; how much worse the model performs without a feature shows us the importance of that feature to our predictions. This procedure is outlined in Figure \ref{process_overview}. Through this procedure, the top 10 most impactful variables become candidates for mobile sensors, leaving 42 unselected variables.


We also performed a forward-stepwise selection process, starting with the 42 base variables that were not selected from the previous LSTM-masking procedure \cite{john1994irrelevant}. This means that we first re-trained our model on the 42 base features and recorded the resulting accuracy. Then, we re-trained our existing model with the 42 base features and one additional feature taken from the top 10 features. This was done for each of the top 10 features, with us calculating and recording the resulting accuracy for each. The feature that resulted in the highest accuracy was then added to the 42 base variables (now giving us a total of 43 variables). For forward-selection based on accuracy, we can continue this process to pick the next best feature by cycling through the features once again to identify which of the nine remaining features produced the highest accuracy. 

Two important notes contextualize this approach. First, the approach is trained on simulated operating data  with a focus on the uncertainty of the faults. Real-world application would require training on data from a specific plant under operating conditions. Second, selecting sensors based on accuracy does not take into account the costs of decisions made as a result of predictions, nor the uncertainty in the predictions. For that, we turn to selecting sensors based on EVSI.



\begin{figure}[t]
\begin{center}

\includegraphics[height=0.87\columnwidth]{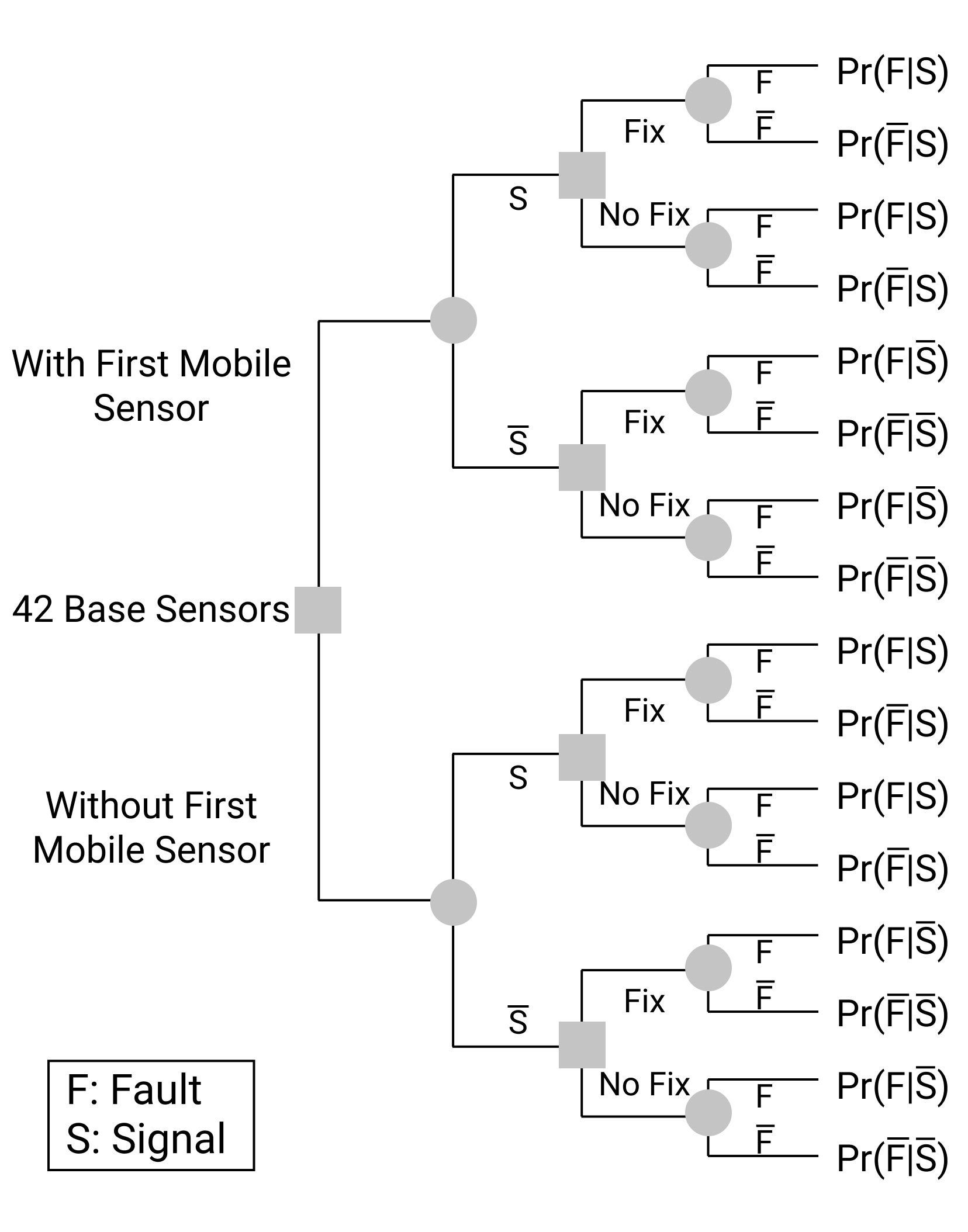}

\end{center}
\caption{Decision Tree for 42 Base Sensors.}
\label{42_sensors}
\end{figure}
\begin{figure}[t]
\begin{center}
\includegraphics[height=0.87\columnwidth]{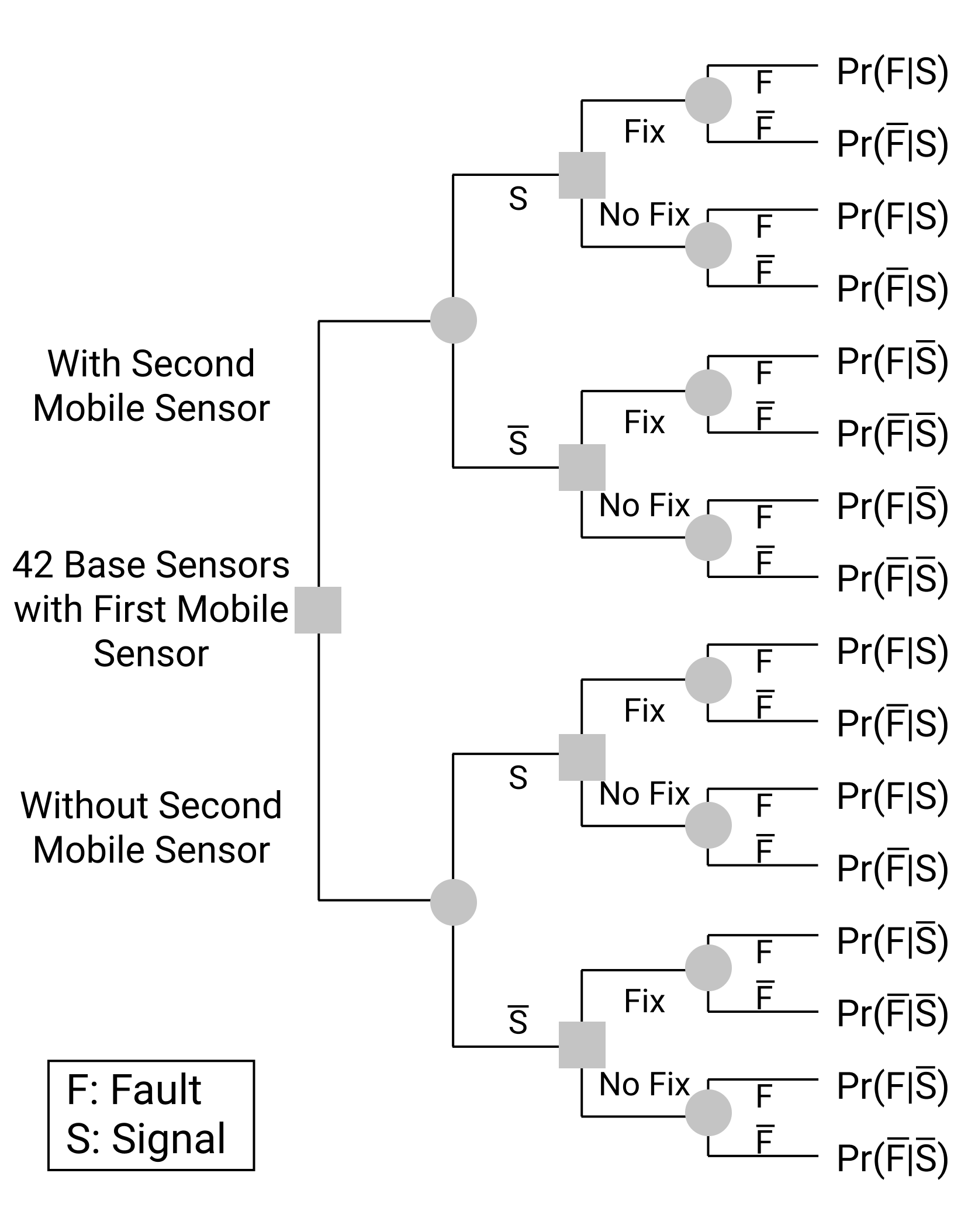}
\end{center}
\caption{Decision Tree for 42 Base Sensors and 1 Mobile Sensor.}
\label{42_sensors_one}
\end{figure}

\subsection*{3.3 Expected Value of Information} 

Expected Value of Information (EVI) is a concept from decision analysis, defined to be the amount a decision maker should be willing to pay for information to reduce or eliminate parametric uncertainty before making critical decisions (see Section 2.3), and is a useful tool for effective management of uncertainty \cite{Applied} \cite{Sensitivity}. Such considerations are crucial for our study, as that selecting sensors based only on accuracy does not account for the costs and benefits of the improved decision making associated with uncertainty reduction with the added sensors.
We assume a simple model to illustrate the framework, where a warning signal is alarmed when triggered by the sensor network as a probable fault in the system.
The conditional probabilities for triggering a warning signal given a fault (or no fault) is taken from the performance of the ML model: $Pr(signal|fault)$, $Pr(\overline{signal}|fault)$,  $Pr(signal|\overline{fault})$, and $Pr(\overline{signal}|\overline{fault})$.  The prior probabilities for faults $Pr(fault)$ and $Pr(\overline{fault})$ are the empirical probabilities observed in the training split from the TE dataset. The probability of triggering a warning signal is:
\begin{align*}\label{eq_Psignal}
    Pr(signal) &=   Pr(signal|fault) * Pr(fault) + \\
    & Pr(signal|\overline{fault})*Pr(\overline{fault}) \tag{1}
\end{align*}

To calculate the conditional probabilities for a fault, Bayes' rule is used.
\begin{align*}\label{eq_Bayes1}
    Pr(fault|signal) &=  \frac{Pr(signal|fault) * Pr(fault)}{Pr(signal)} \tag{2}
\end{align*}
\begin{align*}\label{eq_Bayes2}
    Pr(\overline{fault}|signal) &=  \frac{Pr(signal|\overline{fault}) * Pr(\overline{fault})}{Pr(signal)} \tag{3}
\end{align*}

EVSI is the difference between the expected costs when different sensors are used to send signals\cite{Sensor}.

For illustrative purposes in our application on the sensing TE data set, we simplify the problem by limiting the costs of decisions associated with only true-positive, false-negative and false-positive events. To simplify the presentation, the only costs considered are the cost of remediation (R) and the cost of plant damage if the fault occurs and is not remediated (P). In reality, the cost of placing a certain sensor is dependent on many factors: physical location, type of sensor, method of deployment and time required, etc.  However, since this paper aims to demonstrate a framework, we simplify the sensor placing process by assuming that all sensor deployments have equal costs and are accounted for in the cost for R and P. 

\begin{table}[th!]\label{table:cost}
\vspace{-6mm}
\caption{Costs for EVSI}
\begin{center}
\label{table_ASME}
\begin{tabular}{c l l l}
\hline
Fault Signal & Ground Truth & Action & Cost \\
\hline
$\mathrm{\overline{signal}}$  & $\mathrm{\overline{fault}}$ & no fix &  0 \\
$\mathrm{\overline{signal}}$  & fault & no fix & plant damage (P) \\
signal  & fault & no fix & plant damage (P) \\
signal  & fault / $\mathrm{\overline{fault}}$ & fix & cost of remediation (R)\\
signal  & $\mathrm{\overline{fault}}$ & no fix & 0 \\
\hline
\end{tabular}
\end{center}
\vspace{-7mm}
\end{table}
We assume if remediation is applied to the fault then the only cost is the remediation cost. These situations will be distinguished based on whether or not a fault is actually present or not when a fault warning signal is alarmed. A true-negative, with zero cost associated with it, is the status quo, when there is no warning signal and no fault present. The false-negative case occurs when there is no warning signal alarmed, but there is actually a serious fault that causes loss of the plant’s outcome and damage due to no remediation action. This cost is noted as variable P. If the warning signal is sent and the fault has occurred (true-positive) two actions are possible: (1) remediation with cost R but no damage or (2) No-action, or no fix, because it is believed to be a false positive, resulting in cost P (Figure \ref{42_sensors}).

As there are prior and conditional probabilities associated with these combinations of events (calculated using Equation \ref{eq_Psignal} and \ref{eq_Bayes1}), Bayes rule is applied to the outcomes to determine the expected value of the costs for decision-making. These cases are outlined in Table 1. 

We can calculate the maximum value an expected value decision maker will pay for an additional sensor to be: 
\begin{align*}\label{eq_Cost_with_sensors}
    E(&cost_{sensor}) =  \notag\\ 
    & Pr(signal) *  \min\Big(R*Pr(\overline{fault}|signal),
    P*Pr(fault|signal)\Big)  
    \notag\\ 
     + & Pr(\overline{signal}) * \min\Big(R*Pr(\overline{fault}|\overline{signal}),  P*Pr(fault|\overline{signal})\Big) \tag{4}
\end{align*}

The EVSI of adding the sensor will be the difference between the cost with the sensor and the cost without as shown below:

\begin{align*}\label{evsi_cost_difference}
    EVSI = E(cost_{with\; only\; the\; base\; sensors})-E(&cost_{with\; new\; sensor})  \tag{5}
\end{align*}

The sensor with the highest EVSI is selected to be deployed. We continued this process to pick the next best sensors by cycling through the nine remaining features to identify the sensor that produced the highest EVSI. Therefore, we were able to identify which sensors would best improve EVSI and in what order these sensors should be deployed (Figure \ref{42_sensors_one}). This process continues until adding additional sensors does not improve EVSI or there are no more undeployed sensors. 

\subsection*{3.4 Assumptions} 

Several assumptions are latent in this work. First, the proposed framework assumes trained ML models are available from which to calculate conditional probabilities. These ML models can be trained on data from historical data of a plant, or data collected from a similar plant or a digital twin of a plant. The cost of a sensor deployment was assumed to be the same for all sensors. In practice, some sensors may be more costly to deploy, but the methodology easily allows differential costs into the model.


\begin{figure*}[t!]
\includegraphics[width=\textwidth]{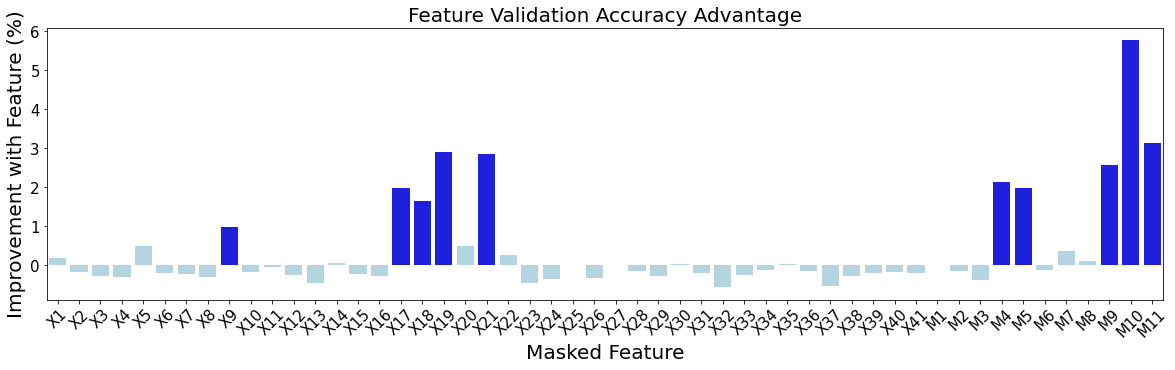}
\caption{\textbf{Masking procedure results for the impact of adding features to the model.} Features with high improvement percentages indicate that they have a high impact on the validation test's accuracy whereas the negative ones lowered the accuracy. The highlighted bars are the top 10 features which we will use for our EVSI analysis. $X_{i}$ variables are measured while $M_{i}$ are manipulated.}
\label{accuracy_advantage}
\end{figure*}

\begin{figure}[t!]
\captionsetup{width=\columnwidth}
\includegraphics[width=\columnwidth]{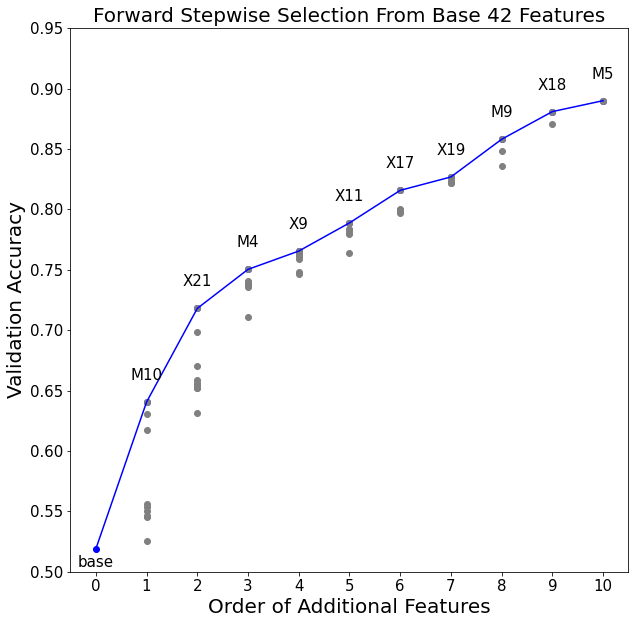}
\caption{\textbf{Incremental increase in validation accuracy} as the 10 highest-impact sensors are added to the base case of 42 features, with the model retrained upon each addition. $X_{i}$ is used to describe measured variables and $M_{i}$ for manipulated.}
\label{stepwise_graph}
\end{figure}

\section*{4. RESULTS}

\subsection*{4.1 Sensors Selected Based on Accuracy}

In Figure \ref{accuracy_advantage}, we show the features and corresponding improvement in validation accuracy  when the feature is included as input to the machine learning model ; the higher the improvement in validation accuracy, the more important that feature.  
We first used the LSTM model to choose the 10 most impactful variables as shown by Figure \ref{accuracy_advantage}. We also performed forward-stepwise selection using these top 10 features, which produced the validation accuracies in Figure \ref{stepwise_graph}. For details about the physical interpretation of each feature, refer to the original TE process paper \cite{downs1993plant}. 

Our base 42 variables had produced an original accuracy of 51.89\% and adding the M10 feature data improved our validation accuracy to 64.05\%, an improvement of almost 5\% over the base variables. The features, as shown by Figure \ref{stepwise_graph},  monotonically increase the validation accuracy. With every additional feature added, the model performs more accurately, suggesting maximizing the number of features could create the best model for our EVSI predictions.
    
    
    
    
    

\begin{table}[th!]
\caption{This table uses the test set to compare each model's precision, recall, and F1-score (eqns. 6-8) by analyzing their weighted averages after training the model on the provided training set of data.}
\begin{center}
\label{ml_comparison}
\begin{tabular}{c | l l l}
\hline
Model & Precision & Recall & F1-Score \\ 
\hline
LSTM & \textbf{0.912} & \textbf{0.898} & \textbf{0.902} \\
CNN & 0.803 & 0.736 & 0.723 \\
Random Forest & 0.909 & 0.877 & 0.886 \\
Logistic & 0.570 & 0.568 & 0.537 \\
\hline
\end{tabular}
\end{center}
\vspace{-10mm}
\end{table}

\subsection*{4.2 Performance Comparison of ML Models}
We then tested the performance of four industry-standard machine learning algorithms on the TE data set: Long Short-Term Memory, Logistic Regression, Random Forest, and 1D Convolutional Neural Network. By using all 52 features and analyzing the predictions on a 20\% validation set, we found that the LSTM model performed the best at predicting the 21 faults considering the metrics of precision, recall, F1 and accuracy (see Appendix A and B).

The validation accuracies produced using multi-label classification are as follows: 89.81\% accuracy for the LSTM model in comparison to 89.0\% for Random Forest, 85.25\% for CNN, and 63.33\% for Logistic Regression. The specifics of the precision and recall are shown in Table \ref{ml_comparison}. Precision (Eqn. 6) is a metric that allows us to determine what percentage of positive results were correct and recall (Eqn. 7) tells us the proportion of actual positives being identified correctly. The F1-score (Eqn. 8) is a single-value indicator to measure accuracy using the harmonic mean of precision and recall. Validation accuracy (Eqn. 9) was used to determine which variables had the highest impact and its equation is shown below. In the equations below, $TP$ is the number of true positives, $TN$ is true negatives, $FP$ is false positives, and $FN$ is false negatives.

\begin{align*}
    \label{PrecisionRecall}
    Precision &= \frac{TP}{TP+FP}  \tag{6}\\
    Recall &= \frac{TP}{TP+FN}  \tag{7}\\
    F_{1} &= \frac{TP}{TP+\frac{1}{2}(FP+FN)} \tag{8}\\
    Accuracy &= \frac{TP+TN}{TP+TN+FP+FN} \tag{9}
\end{align*}

For each model, we tuned the hyperparameters to best model and predict the data. While the logistic regression and CNN algorithms produced significantly lower validation accuracies, we can observe that the LSTM is the best algorithm among the four for chemical plant data as per the TE data set. Confusion matrices for CNN, Random Forest, and Logistic Regression approaches are shown in Appendix B for comparison.

\subsection*{4.3 Ranking of Sensor Selection Based on EVSI and  Accuracy Improvement} 
Table \ref{ESVI_select} shows the EVSI calculated to select the next highest-priority sensor for monitoring. The first two columns show the sensors that provides the greatest value if monitored out of the 10 most impactful sensors. Similarly, the second column shows which sensor is the next most useful sensor to monitor, if we are already monitoring the most impactful sensor out of the 10 sensors (M10, in this case). 

Note that we rerun the EVSI process again after we added M10 because we assume that there could be correlations between features. We observe that sensor EVSI rankings did not remain constant as sensors were added: in Table 3, for example, while X9 is the second-highest ranked sensor during the first selection, after M10 is chosen (second column), it is not the highest-ranked sensor, but the third-highest ranked. Table \ref{fs_select} shows the accuracy improvement when we add sensor data to the LSTM model that was originally trained with 42 base sensors and retrain it.

\begin{table}[th]
\caption{EVSI for selecting the first three sensors. Sensors ranked based on EVSI calculated in descending order.}
\label{ESVI_select}
\scriptsize
\begin{tabular}{c  c | c c| c c}
\hline
Sensor & EVSI, 1st Sensor & Sensor & EVSI, 2nd Sensor & Sensor & EVSI, 3rd Sensor\\
\hline
\textbf{M10} & \textbf{0.59} & \textbf{X21} & \textbf{0.29} & \textbf{M5} & \textbf{0.15}\\
X9 & 0.51 & M5 & 0.20 & X18 & 0.07\\
X21 & 0.47 & X9 & 0.18 & M4 & 0.0\\
M5 & 0.26 & M11 & 0.10 & X19 & 0.04\\
M9 & 0.15 & M4 & 0.06 & X17 & -0.06\\
X18 & 0.10 & X18 & 0.06 & M9 & -0.08\\
X17 & 0.05 & M9 & 0.05 & M11 & -0.09\\
M11 & 0.0 & X17 & 0.04 & X9 & -0.10\\
X19 & 0.0 & X19 & 0.02 & N/A & N/A\\
M4 & 0.0 & N/A & N/A & N/A & N/A\\

\hline
\end{tabular}
\vspace{-4mm}
\end{table}

\begin{table}[th]
\centering 
\caption{Accuracy improvement of the LSTM model by adding one sensor data to the training set. Sensor ranked in descending order.}
\label{fs_select}
\scriptsize
\begin{tabular}{c  c | c c}
\hline
Sensor & Acc Improve, 1st Sensor & Sensor & Acc Improve, 2nd Sensor \\
\hline
\textbf{M10} & \pmb{$12\%$} & \textbf{X21} & \pmb{$8\%$}\\
X21 & $11\%$ & X9 & $6\%$\\
X9 & $10\%$ & M4 & $3\%$\\
M4 & $4\%$ & X18 & $2\%$\\
X18 & $3\%$ & M11 & $2\%$\\
M9 & $3\%$ & X19 & $2\%$\\
M5 & $3\%$ & M5 & $1\%$\\
M11 & $3\%$ & M9 & $1\%$\\
X19 & $3\%$ & X17 & $-0.7\%$\\
X17 & $0.7\%$ & N/A & N/A\\

\hline
\end{tabular}
\vspace{-4mm}
\end{table}

\subsection*{4.4 
Sensitivity to Cost Ratio (P/R)} 

Table \ref{cost_sensitivity} shows the different EVSI values calculated on the same set of sensors over different ratios of costs of damage to the plant (P) to the cost of remediation (R). The cost ratios are calculated as $\frac{P}{R}$ and are all powers of $2$. We observe that changing this ratio does not change the relative ordering of the top three sensors. The remaining sensors in the top ten shift ranks relatively, but never move into the top three. We note that among the trailing seven sensors, the changes in relative ranks are not substantial across different ratios. 

Besides the EVSI score, Table \ref{cost_sensitivity} also shows the best action (based on EVSI) taken when a new sensor is added. For example, in the first row of the top left section (P/R=2), "Fix" under Action (Signal) means that if a robot is deployed with sensor M10, then the optimal action would be "Fix" when a fault signal is observed. Similarly, we would take the action of "No Fix" when no fault signal is observed. Because our machine learning model that represents a sensor in this case that has very high accuracy, the EVSI decision process follows the signal indication; the remediation is only taken if the signal is present. This trend continues until the EVSI goes to zero with high ratios of P/R, where the EVSI also goes to zero.

\begin{table*}[h!]
\centering
    \scriptsize
\caption{EVSI of 10 Sensors with Different Ratios of Cost of P and R (Ranked in Descending Order)}
\label{cost_sensitivity}
\begin{tabular}{c  c  c  c| c c  c c }
\hline
Sensor& EVSI (P/R = $2^1$) & Action (Signal) & Action (No Signal) & Sensor& EVSI (P/R = $2^2$) & Action (Signal) & Action (No Signal) \\
\hline
\textbf{M10} & 0.24 & Fix & No Fix & \textbf{M10} & 0.45 & Fix & No Fix  \\
\textbf{X9} & 0.17 & Fix & No Fix & \textbf{X9} & 0.37 & Fix & No Fix \\
\textbf{X21} & 0.14 & Fix & No Fix & \textbf{X21} & 0.34 & Fix & No Fix  \\
X18 & 0.06 & Fix & No Fix & M5 & 0.21 & Fix & No Fix \\
M9 & 0.05 & Fix & No Fix & M9 & 0.17 & Fix & No Fix \\
M5 & 0.05 & Fix & No Fix & X18 & 0.16 & Fix & No Fix \\
X19 & 0.04 & Fix & No Fix & X17 & 0.12 & Fix & No Fix \\
X17 & 0.02 & Fix & No Fix & M4 & 0.10 & Fix & No Fix \\
M4 & 0.02 & Fix & No Fix & M11 & 0.10 & Fix & No Fix \\
M11 & 0.01 & Fix & No Fix & X19 & 0.08 & Fix & No Fix \\

\hline
\end{tabular}
\begin{tabular}{c  c c  c| c c  c c }
\hline
 Sensor & EVSI (P/R = $2^3$) & Action (Signal) & Action (No Signal) &  Sensor & EVSI(P/R = $2^4$) & Action (Signal) & Action (No Signal)\\
\hline
\textbf{M10} & 0.59 & Fix & No Fix & \textbf{M10} & 0.0 & Fix & Fix\\
\textbf{X9} & 0.51 & Fix & No Fix & \textbf{X9} & 0.0 & Fix & Fix \\
\textbf{X21} & 0.47 & Fix & No Fix & \textbf{X21} & 0.0 & Fix & Fix \\
M5 & 0.26 & Fix & No Fix & M5 & 0.0 & Fix & Fix\\
M9 & 0.15 & Fix & No Fix & M9 & 0.0 & Fix & Fix\\
X18 & 0.10 & Fix & No Fix & X18 & 0.0 & Fix & Fix\\
X17 & 0.05 & Fix & No Fix & X17 & 0.0 & Fix & Fix\\
M11 & 0.0 & Fix & Fix & M11 & 0.0 & Fix & Fix\\
X19 & 0.0 & Fix & Fix & X19 & 0.0 & Fix & Fix\\
M4 & 0.0 & Fix & Fix & M4 & 0.0 & Fix & Fix\\

\hline
\end{tabular}
\end{table*}

\section*{5. DISCUSSION}
\subsection*{5.1 Top Candidate Sensors}
We use the order of adding features based on the values in Figure \ref{stepwise_graph} and ESVI in the forward stepwise-selection process. In our example, M10 shows a high validation accuracy increase, which means access to that feature will increase model performance. For the TE dataset, we conclude that an initial pair of sensors (M10 and X21) added yield an outsize proportion of validation accuracy improvement (an improvement of 0.2, from 0.52 to 0.72) across the top ten most impactful sensors identified. The incremental validation accuracy provided by the first pair of sensors is notable, given that with all 52 features engaged, a validation accuracy of roughly 0.90 is achieved. 

While these findings are specific to the TE data, they illustrate how EVSI can be used to efficiently deploy sensors in order to reduce the uncertainty in detecting faults in a chemical plant. Machine learning was used to determine conditional probabilities and estimate the expected value for an additional sensor deployment to inform plant operators about recommended decisions. With a relatively small set of sensors incrementally added to a baseline of sensor data, we can produce a relatively high validation accuracy under conditions where several sensors are not functioning. These findings suggest that during plant emergency response, machine learning models can help resolve uncertainty in making decisions about where and toward what purpose to deploy sensing platforms.

\subsection*{5.2 Sensitivity Analysis on the Cost Function}

We observe that if the ratio of $P/R$ is low enough, the EVSI process  will always advise "No Fix" when there is no fault signal: fixing the fault is costlier than the expected cost of incurring plant damage (Table 5). However, as the ratio of $P/R$ reaches 16, the EVSI decision process will always advise "Fix", because the cost of the plant failing outweighs remediation costs and there would be no risk that the "no signal" condition is a false negative. In practice, this would result in a  continuous preventive maintenance program.

Similarly, if we look at the lower-priority sensors (M11, X19, M4) for lower ratios (e.g., P/R=8), we see the same behavior of always to "Fix" even when there is no fault signal. From Table \ref{fs_select}, we see these lower-priority sensors are mostly sensors that bring limited accuracy increase to the machine learning model; if a robot is sent to deploy these sensors, the fault signal received is not very accurate and may have a high false positive or false negative rate. At the P/R ratio of 8, with these lower-priority sensors  the EVSI decision process is already starting to devalue the expected value of a new sensor due to higher uncertainty in the results.

These results highlight the interplay between the $P/R$ cost ratio and accuracy of sensors, emphasizing the potential of rapidly-deployable sensing platforms to afford decision-making facilitated by an EVSI-based approach. 


\begin{figure}[t]
\includegraphics[width=\columnwidth]{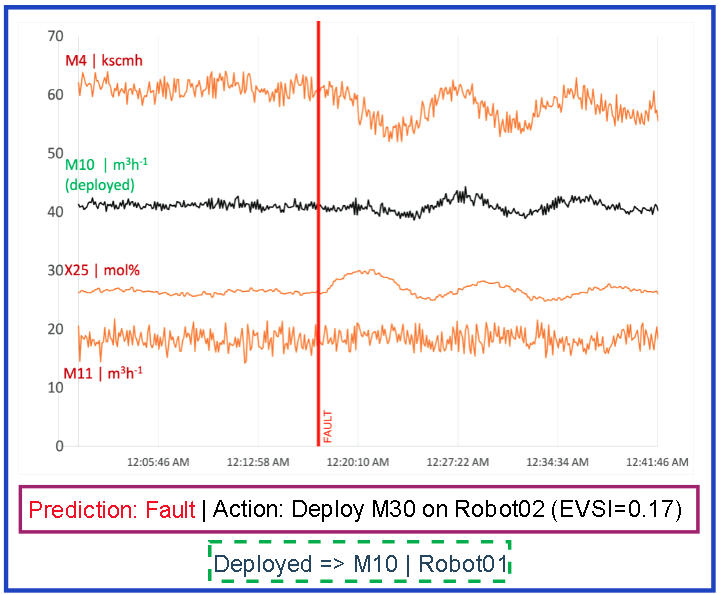}
\caption{A dashboard with data from the PI System that displays data from selected critical sensors, results of the initial EVSI analysis and recommends which sensors should be deployed on mobile robots.}
\label{osi_interface}
\end{figure}

\subsection*{5.3 Implications for Research and Practice.}

This paper provides a framework for applying EVSI for real time mobile sensing applications in industrial environments. We argue that EVSI can be utilized in any situation that calls for informed decision-making, especially in the presence of faulty or sparse data. Several assumptions, outlined in Section 3.4, present rich opportunities for future research in applying ML to manage uncertainty and risk response in complex plant environments.

To envision how such a system could create impact in practice, Fig. \ref{osi_interface} shows a sample user interface and scenario from an operator's standpoint using the PI\texttrademark{} System \cite{osisoft_2021}, a leading industrial operations data management platform. In this scenario a fault is predicted  at 12:17:30 AM, and, leveraging our proposed framework, EVSI results recommended deploying sensor M10 on mobile robot Robot01 to the site. The system calculates the EVSI for all remaining un-deployed sensors and recommends the operator deploy M30, the sensor with the maximum EVSI next. Considering incoming data from the newly-deployed robot, an operator could make more informed decisions and evaluate risks. Finally, while the findings in this work are validated using the TE dataset, we anticipate that this overall workflow can be generalized to emergency sensing and actuation decision-making in a variety of plant contexts. 

\section*{6. CONCLUSIONS AND FUTURE RESEARCH}

This paper presents an approach to sensor deployment in response to fault prediction using Expected Value of Sample Information (EVSI). Conditional probabilities were learned over a data set derived from the Tennessee Eastman plant simulation dataset. We are able to illustrate that (1) the LSTM model was the most accurate to describe the TE dataset, (2) the model indicated which ten sensors had most influence on accuracy, and (3) the deployment framework is appropriately sensitive to plant damage and remediation costs.   

Further work aims to consider the risks, costs and time associated with  sensor deployment approaches (e.g., human-installation, drops from drones, ground robots, throw-bots, rotory robots) to achieve optimal sensor placements. This research will also consider additional signals or virtual sensors to trigger an EVSI evaluation. Furthermore, the TE Process data lack information on cost, location, time, and accessibility of various sensors in the simulated plant environment – all information that is critical to decision-making in sensor deployment. Extending our approach to real data would allow consideration of these factors and result in a more impactful analysis.

\section*{ACKNOWLEDGEMENTS}

The authors would like to acknowledge Douglas Hutchings and Devin Pon for their technical assistance on IT support. We value our community partnership with OSIsoft (now part of AVEVA), maker of the PI System, in developing this case study.




%
% ...

\bibliographystyle{asmems4}
\bibliography{main}

\begin{thebibliography}{10}

\bibitem{kulkarni2005knowledge}
Kulkarni, A., Jayaraman, V.~K., and Kulkarni, B.~D., 2005.
\newblock ``{Knowledge Incorporated Support Vector Machines to Detect Faults in
  Tennessee Eastman Process}''.
\newblock {\em Computers and Chemical Engineering, \textbf{ 29}}(10),
  pp.~2128--2133.

\bibitem{10.1115/IMECE2017-71829}
Bergonzi, L., Colombo, G., Rossoni, M., and Furini, F., 2017.
\newblock ``Data and {Knowledge} in {IIoT-Based} {Maintenance} {Application}''.
\newblock Vol.~11: Systems, Design, and Complexity of {\em ASME International
  Mechanical Engineering Congress and Exposition}.

\bibitem{10.1115/IMECE2018-88262}
Alothman, H.~A., Khasawneh, M.~T., and Nagarur, N.~N., 2018.
\newblock ``{Internet of Things in Manufacturing: An Overview}''.
\newblock Vol.~Volume 2: Advanced Manufacturing of {\em ASME International
  Mechanical Engineering Congress and Exposition}.
\newblock V002T02A076.

\bibitem{4796311}
{Gungor}, V.~C., and {Hancke}, G.~P., 2009.
\newblock ``{Industrial Wireless Sensor Networks: Challenges, Design
  Principles, and Technical Approaches}''.
\newblock {\em IEEE Transactions on Industrial Electronics, \textbf{ 56}}(10),
  pp.~4258--4265.

\bibitem{10.1115/1.2951943}
Xinmin, L., Zhaoqing, T., and Zhongqin, L., 2008.
\newblock ``{A Simplified Method for Optimal Sensor Distribution for Process
  Fault Diagnosis in Multistation Assembly Processes}''.
\newblock {\em Journal of Manufacturing Science and Engineering, \textbf{
  130}}(5), 08.
\newblock 051002.

\bibitem{10.1115/1.2830221}
Khan, A., Ceglarek, D., and Ni, J., 1998.
\newblock ``{Sensor Location Optimization for Fault Diagnosis in Multi-Fixture
  Assembly Systems}''.
\newblock {\em Journal of Manufacturing Science and Engineering, \textbf{
  120}}(4), 11, pp.~781--792.

\bibitem{savazzi2013wireless}
Savazzi, S., Guardiano, S., and Spagnolini, U., 2013.
\newblock ``{Wireless Sensor Network Modeling and Deployment Challenges in Oil
  and Gas Refinery Plants}''.
\newblock {\em International Journal of Distributed Sensor Networks, \textbf{
  9}}(3), p.~383168.

\bibitem{7588228}
{Lin}, C., {Deng}, D., {Chen}, Z., and {Chen}, K., 2016.
\newblock ``{Key Design of Driving Industry 4.0: Joint Energy-Efficient
  Deployment and Scheduling in Group-Based Industrial Wireless Sensor
  Networks}''.
\newblock {\em IEEE Communications Magazine, \textbf{ 54}}(10), pp.~46--52.

\bibitem{wang2016}
Wang, R., Veloso, M., and Seshan, S., 2016.
\newblock ``{Active Sensing Data Collection with Autonomous Mobile Robots}''.
\newblock In 2016 IEEE International Conference on Robotics and Automation
  (ICRA), pp.~2583--2588.

\bibitem{SquishyWeb}
Squishy-Robotics.
\newblock \url{https://squishy-robotics.com/}, {Accessed: 2021-04-20}.

\bibitem{downs1993plant}
Downs, J.~J., and Vogel, E.~F., 1993.
\newblock ``{A Plant-Wide Industrial Process Control Problem}''.
\newblock {\em Computers and Chemical Engineering, \textbf{ 17}}(3),
  pp.~245--255.

\bibitem{TEextended}
Rieth, C.~A., Amsel, B.~D., Tran, R., and Cook, M.~B.
\newblock {Additional Tennessee Eastman Process Simulation Data for Anomaly
  Detection Evaluation}, {2017}, {V1}, {https://doi.org/10.7910/DVN/6C3JR1}
  {Accessed: 2021-04-20}.

\bibitem{BALLARI2012102}
Ballari, D., de~Bruin, S., and Bregt, A., 2012.
\newblock ``{Value of Information and Mobility Constraints for Sampling with
  Mobile Sensors}''.
\newblock {\em Computers and Geosciences, \textbf{ 49}}, pp.~102--111.

\bibitem{schmidt_smith_hite_mattingly_azmy_rajan_goldhahn_2019}
Schmidt, K., Smith, R.~C., Hite, J., Mattingly, J., Azmy, Y., Rajan, D., and
  Goldhahn, R., 2019.
\newblock ``{Sequential Optimal Positioning of Mobile Sensors Using Mutual
  Information}''.
\newblock {\em STATISTICAL ANALYSIS AND DATA MINING}.

\bibitem{8968086}
Cera, B.~M., Thompson, A., and Agogino, A.~M., 2019.
\newblock ``{Energy-Efficient Locomotion Strategies and Performance Benchmarks
  using Point Mass Tensegrity Dynamics}''.
\newblock In 2019 IEEE/RSJ International Conference on Intelligent Robots and
  Systems (IROS), pp.~4678--4683.

\bibitem{6387986}
{Wang}, H., {Yao}, K., and {Estrin}, D., 2005.
\newblock ``{Information-Theoretic Approaches for Sensor Selection and
  Placement in Sensor Networks for Target Localization and Tracking}''.
\newblock {\em Journal of Communications and Networks, \textbf{ 7}}(4),
  pp.~438--449.

\bibitem{7487415}
Wang, R., Veloso, M., and Seshan, S., 2016.
\newblock ``{Active Sensing Data Collection with Autonomous Mobile Robots}''.
\newblock In 2016 IEEE International Conference on Robotics and Automation
  (ICRA), pp.~2583--2588.

\bibitem{933092}
Romero, L., Morales, E., and Sucar, E., 2001.
\newblock ``{An Exploration and Navigation Approach for Indoor Mobile Robots
  Considering Sensor's Perceptual Limitations}''.
\newblock In Proceedings 2001 ICRA. IEEE International Conference on Robotics
  and Automation (Cat. No.01CH37164), Vol.~3, pp.~3092--3097 vol.3.

\bibitem{10.1115/1.4036014}
Chen, L.-H., Kim, K., Tang, E., Li, K., House, R., Zhu, E.~L., Fountain, K.,
  Agogino, A.~M., Agogino, A., Sunspiral, V., and Jung, E., 2017.
\newblock ``{Soft Spherical Tensegrity Robot Design Using Rod-Centered
  Actuation and Control}''.
\newblock {\em Journal of Mechanisms and Robotics, \textbf{ 9}}(2), 03.

\bibitem{Luo2018TensegrityRL}
Luo, J., Edmunds, R., Rice, F., and Agogino, A., 2018.
\newblock ``{Tensegrity Robot Locomotion under Limited Sensory Inputs via Deep
  Reinforcement Learning}''.
\newblock {\em 2018 IEEE International Conference on Robotics and Automation
  (ICRA)}, pp.~6260--6267.

\bibitem{he2007fault}
He, Q.~P., and Wang, J., 2007.
\newblock ``{Fault Detection Using the K-Nearest Neighbor Rule for
  Semiconductor Manufacturing Processes}''.
\newblock {\em IEEE Transactions on Semiconductor mMnufacturing, \textbf{
  20}}(4), pp.~345--354.

\bibitem{venkatasubramanian2003review}
Venkatasubramanian, V., Rengaswamy, R., Yin, K., and Kavuri, S.~N., 2003.
\newblock ``{A Review of Process Fault Detection and Diagnosis: Part I:
  Quantitative Model-Based Methods}''.
\newblock {\em Computers and Chemical Engineering, \textbf{ 27}}(3),
  pp.~293--311.

\bibitem{park2020review}
Park, Y.-J., Fan, S.-K.~S., and Hsu, C.-Y., 2020.
\newblock ``{A Review on Fault Detection and Process Diagnostics in Industrial
  Processes}''.
\newblock {\em Processes, \textbf{ 8}}(9), p.~1123.

\bibitem{hu2016machine}
Hu, R.~L., 2016.
\newblock ``{Machine Learning to Scale Fault Detection in Smart Energy
  Generation and Building Systems}''.
\newblock PhD thesis, UC Berkeley.

\bibitem{HU2019117}
Hu, R., Granderson, J., Auslander, D., and Agogino, A., 2019.
\newblock ``{Design of Machine Learning Models With Domain Experts for
  Automated Sensor Selection for Energy Fault Detection}''.
\newblock {\em Applied Energy, \textbf{ 235}}, pp.~117--128.

\bibitem{UDUGAMA202020}
Udugama, I.~A., Gernaey, K.~V., Taube, M.~A., and Bayer, C., 2020.
\newblock ``{A Novel Use for an Old Problem: The Tennessee Eastman Challenge
  Process as an Activating Teaching Tool}''.
\newblock {\em Education for Chemical Engineers, \textbf{ 30}}, pp.~20--31.

\bibitem{TEData2019}
Chen, X., 2019.
\newblock {Tennessee Eastman Simulation Dataset}.
\newblock https://dx.doi.org/10.21227/4519-z502, {IEEE Dataport}.

\bibitem{yin2014study}
Yin, S., Gao, X., Karimi, H.~R., and Zhu, X., 2014.
\newblock ``{Study on Support Vector Machine-Based Fault Detection in Tennessee
  Eastman Process}''.
\newblock In Abstract and Applied Analysis, Vol.~2014, Hindawi Journal.

\bibitem{verron2008fault}
Verron, S., Tiplica, T., and Kobi, A., 2008.
\newblock ``{Fault Detection and Identification with a New Feature Selection
  Based on Mutual Information}''.
\newblock {\em Journal of Process Control, \textbf{ 18}}(5), pp.~479--490.

\bibitem{chiang2004fault}
Chiang, L.~H., Kotanchek, M.~E., and Kordon, A.~K., 2004.
\newblock ``{Fault Diagnosis Based on Fisher Discriminant analysis and Support
  Vector Machines}''.
\newblock {\em Computers and Chemical Engineering, \textbf{ 28}}(8),
  pp.~1389--1401.

\bibitem{9444425}
Akbari, M., and Khoshnood, A., 2021.
\newblock ``{A New Feature Selection-Aided Observer for Sensor Fault Diagnosis
  of an Industrial Gas Turbine}''.
\newblock {\em IEEE Sensors Journal}.

\bibitem{7324139}
Ramya, R.~S., and Kumaresan, S., 2015.
\newblock ``{Analysis of Feature Selection Techniques in Credit Risk
  Assessment}''.
\newblock In 2015 International Conference on Advanced Computing and
  Communication Systems, pp.~1--6.

\bibitem{cai2018feature}
Cai, J., Luo, J., Wang, S., and Yang, S., 2018.
\newblock ``{Feature Selection in Machine Learning: A New Perspective}''.
\newblock {\em Neurocomputing, \textbf{ 300}}, pp.~70--79.

\bibitem{https://doi.org/10.1111/j.1539-6924.1999.tb00395.x}
Hammitt, J.~K., and Shlyakhter, A.~I., 1999.
\newblock ``{The Expected Value of Information and the Probability of
  Surprise}''.
\newblock {\em Risk Analysis, \textbf{ 19}}(1), pp.~135--152.

\bibitem{ijcai2017-215}
Choi, Y., Darwiche, A., and den Broeck, G.~V., 2017.
\newblock ``{Optimal Feature Selection for Decision Robustness in Bayesian
  Networks}''.
\newblock In Proceedings of the Twenty-Sixth International Joint Conference on
  Artificial Intelligence, {IJCAI-17}, pp.~1554--1560.

\bibitem{doi:10.1177/0272989X07302555}
Brennan, A., Kharroubi, S., O'Hagan, A., and Chilcott, J., 2007.
\newblock ``{Calculating Partial Expected Value of Perfect Information via
  Monte Carlo Sampling Algorithms}''.
\newblock {\em Medical Decision Making, \textbf{ 27}}(4), pp.~448--470.
\newblock PMID: 17761960.

\bibitem{doi:10.1177/1740774508098413}
Willan, A.~R., 2008.
\newblock ``{Optimal Sample Size Determinations from an Industry Perspective
  Based on the Expected Value of Information}''.
\newblock {\em Clinical Trials, \textbf{ 5}}(6), pp.~587--594.
\newblock PMID: 19029207.

\bibitem{doi:10.1177/0272989X09344746}
Griffin, S., Welton, N.~J., and Claxton, K., 2010.
\newblock ``{Exploring the Research Decision Space: The Expected Value of
  Information for Sequential Research Designs}''.
\newblock {\em Medical Decision Making, \textbf{ 30}}(2), pp.~155--162.
\newblock PMID: 20040743.

\bibitem{Sensor}
de~Bruin, S., Ballari, D., and Bregt, A., 2012.
\newblock ``{Multiphase Sensor Placement Using Expected Value of
  Information}''.
\newblock {\em International Dairy Journal - INT DAIRY J}, 01.

\bibitem{doi:10.1287/ijoc.1090.0327}
{Chick, Stephen E. and Branke, Jürgen and Schmidt, Christian}, 2010.
\newblock ``{Sequential Sampling to Myopically Maximize the Expected Value of
  Information}''.
\newblock {\em INFORMS Journal on Computing, \textbf{ 22}}(1), pp.~71--80.

\bibitem{Memarzadeh2016ValueOI}
Memarzadeh, M., and Pozzi, M., 2016.
\newblock ``{Value of Information in Sequential Decision Making: Component
  Inspection, Permanent Monitoring and System-Level Scheduling}''.
\newblock {\em Reliab. Eng. Syst. Saf., \textbf{ 154}}, pp.~137--151.

\bibitem{MALINGS2016219}
Malings, C., and Pozzi, M., 2016.
\newblock ``{Value of Information for Spatially Distributed Systems:
  Application to Sensor Placement}''.
\newblock {\em Reliability Engineering and System Safety, \textbf{ 154}},
  pp.~219--233.

\bibitem{hochreiter1997long}
Hochreiter, S., and Schmidhuber, J., 1997.
\newblock ``{Long Short-Term Memory}''.
\newblock {\em Neural computation, \textbf{ 9}}(8), pp.~1735--1780.

\bibitem{fukushima1982neocognitron}
Fukushima, K., and Miyake, S., 1982.
\newblock ``{Neocognitron: A Self-Organizing Neural Network Model for a
  Mechanism of Visual Pattern Recognition}''.
\newblock In {\em Competition and cooperation in neural nets}. Springer,
  pp.~267--285.

\bibitem{cox1958regression}
Cox, D.~R., 1958.
\newblock ``{The Regression Analysis of Binary Sequences}''.
\newblock {\em Journal of the Royal Statistical Society: Series B
  (Methodological), \textbf{ 20}}(2), pp.~215--232.

\bibitem{breiman2001random}
Breiman, L., 2001.
\newblock ``{Random Forests}''.
\newblock {\em Machine learning, \textbf{ 45}}(1), pp.~5--32.

\bibitem{john1994irrelevant}
John, G.~H., Kohavi, R., and Pfleger, K., 1994.
\newblock ``{Irrelevant Features and the Subset Selection Problem}''.
\newblock In {\em Machine Learning Proceedings 1994}. Elsevier, pp.~121--129.

\bibitem{Applied}
Raiffa, H., and Schlaifer, R., 1961.
\newblock {\em Applied Statistical Decision Theory}.
\newblock Harvard University Press, Boston, MA.

\bibitem{Sensitivity}
Felli, J.~C., and Hazen, G.~B., 1998.
\newblock ``{Sensitivity Analysis and the Expected Value of Perfect
  Information}''.
\newblock {\em Medical Decision Making, \textbf{ 18}}, pp.~95--–109.

\bibitem{osisoft_2021}
\uppercase {AVEVA-OSI}soft.
\newblock {PI\texttrademark{} System: The operations data management platform
  that engineers and CIOs trust.}
\newblock \url{https://www.aveva.com/en/products/pi-system/}.
\newblock Accessed: 2021-04-20.

\bibitem{kohavi1995study}
Kohavi, R., et~al., 1995.
\newblock ``{A Study of Cross-Validation and Bootstrap for Accuracy Estimation
  and Model Selection}''.
\newblock In Ijcai, Vol.~14, Montreal, Canada, pp.~1137--1145.

\bibitem{abadi2016tensorflow}
Abadi, M., Barham, P., Chen, J., Chen, Z., Davis, A., Dean, J., Devin, M.,
  Ghemawat, S., Irving, G., Isard, M., et~al., 2016.
\newblock ``{Tensorflow: A System for Large-Scale Machine Learning}''.
\newblock In 12th USENIX Symposium on Operating Systems Design and
  Implementation, pp.~265--283.

\bibitem{sawant_2019}
Sawant, M., 2019.
\newblock {Tennessee Eastman Process Simulation Data for Anomaly Detection
  Evaluation}, Jul.

\end{thebibliography}

\newpage
\appendix       
\section*{Appendix A: Machine Learning Model Details}
Using 5-fold cross-validation \cite{kohavi1995study}, we tuned the hyperparameters of the logistic regression and random forest models. The logistic regression model requires one hyperparameter C, the inverse of regularization strength, which we chose to have  a value of 1000 through cross validation on the TE data training set. The random forest model used cross-validation to tune the hyperparameters of 500 trees (estimators) in the random forest, no maximum depth, and the minimum number of samples required to split an internal node set to five. The 1D CNN and LSTM models were implemented in Tensorflow Keras \cite{abadi2016tensorflow}. We added Dropout and trained the models using the Adam optimizer due to stochastic gradient descent handling sparse gradients. We used softmax for activation as there are multiple classes and we calculated loss via categorical cross-entropy.

\section*{Appendix B: Machine Learning Confusion Matrices}

\begin{figure}[h!]
\centering
\includegraphics[width=\columnwidth]{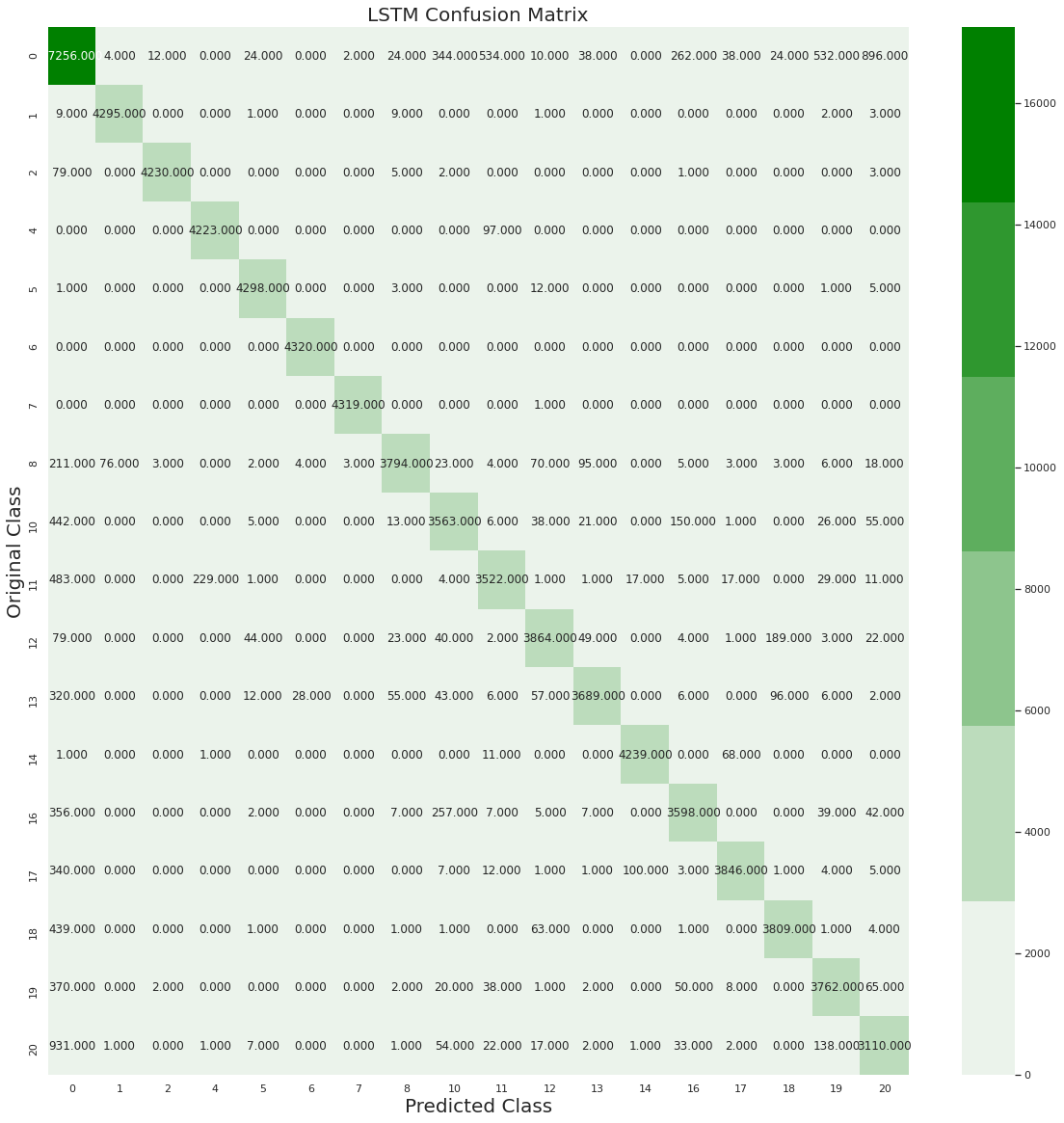}
\caption{LSTM confusion matrix}
\label{confusion_matrix}
\end{figure}

\begin{figure}[h!]
\centering
\includegraphics[width=\columnwidth]{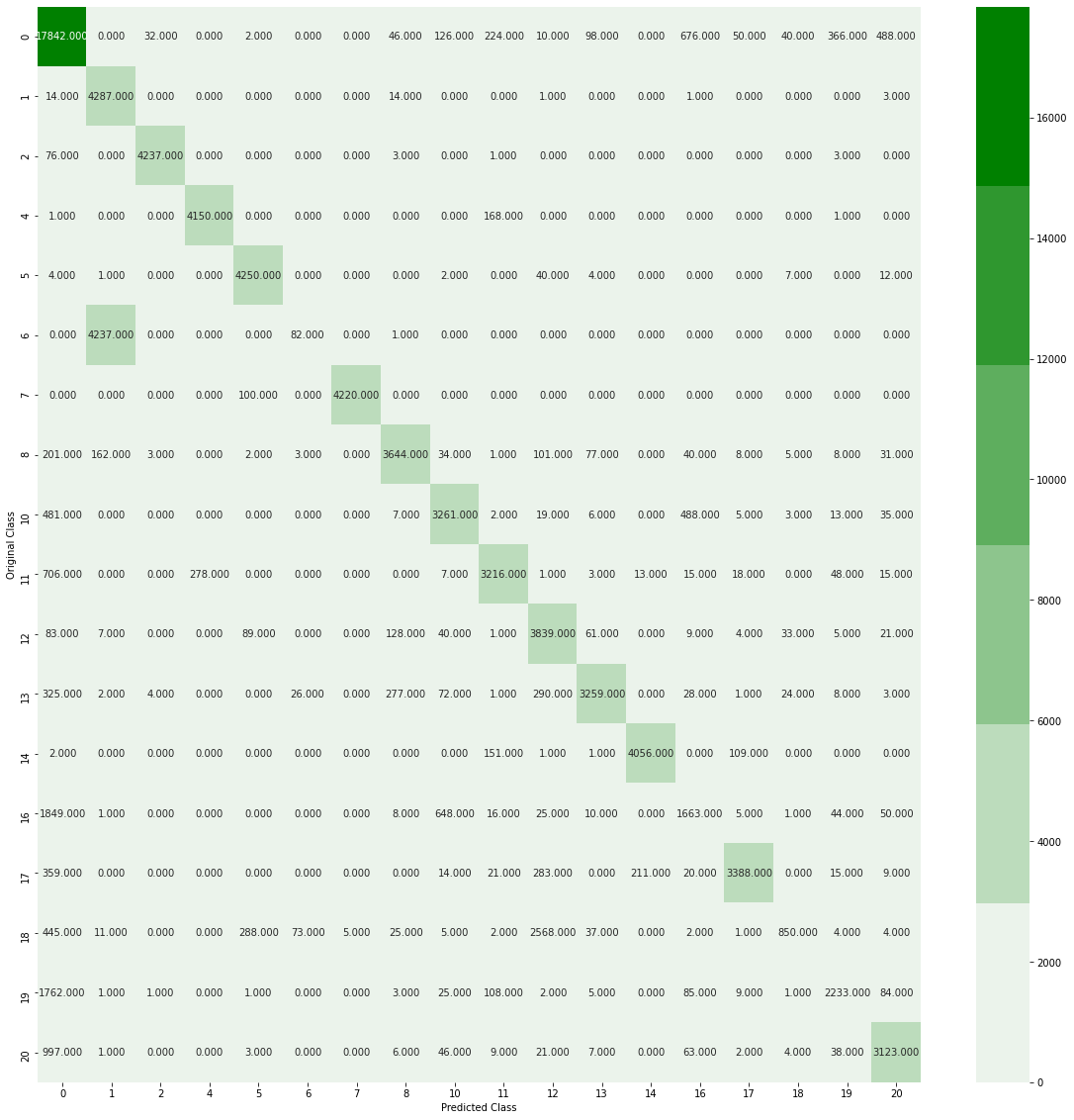}
\caption{1D Convolution Neural Network confusion matrix}
\label{cnn_confusion_matrix}
\end{figure}

\begin{figure}[h!]
\centering
\includegraphics[width=\columnwidth]{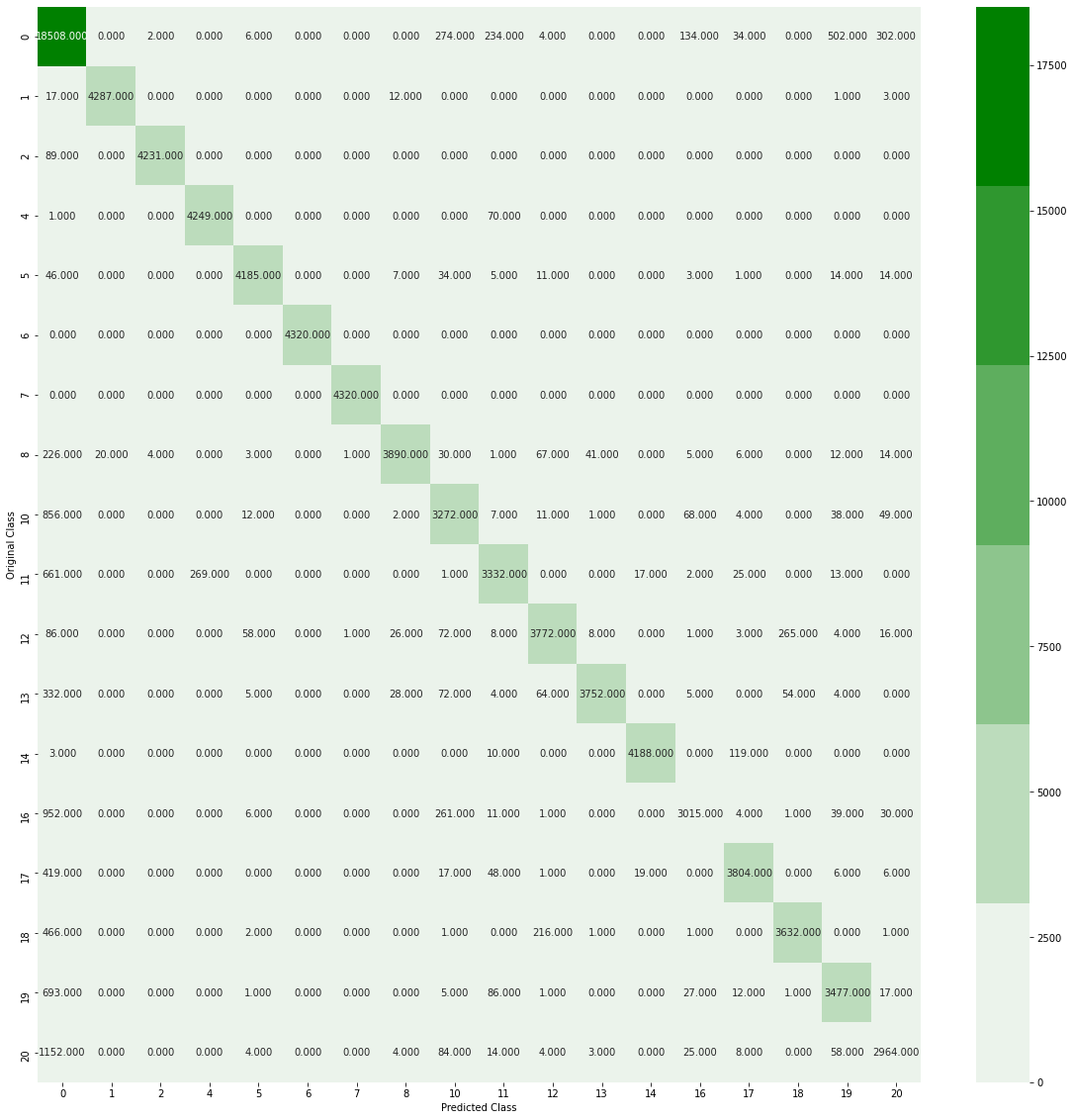}
\caption{Random Forest confusion matrix}
\label{randomforest_confusion_matrix}
\end{figure}

\begin{figure}[h!]
\centering
\includegraphics[width=\columnwidth]{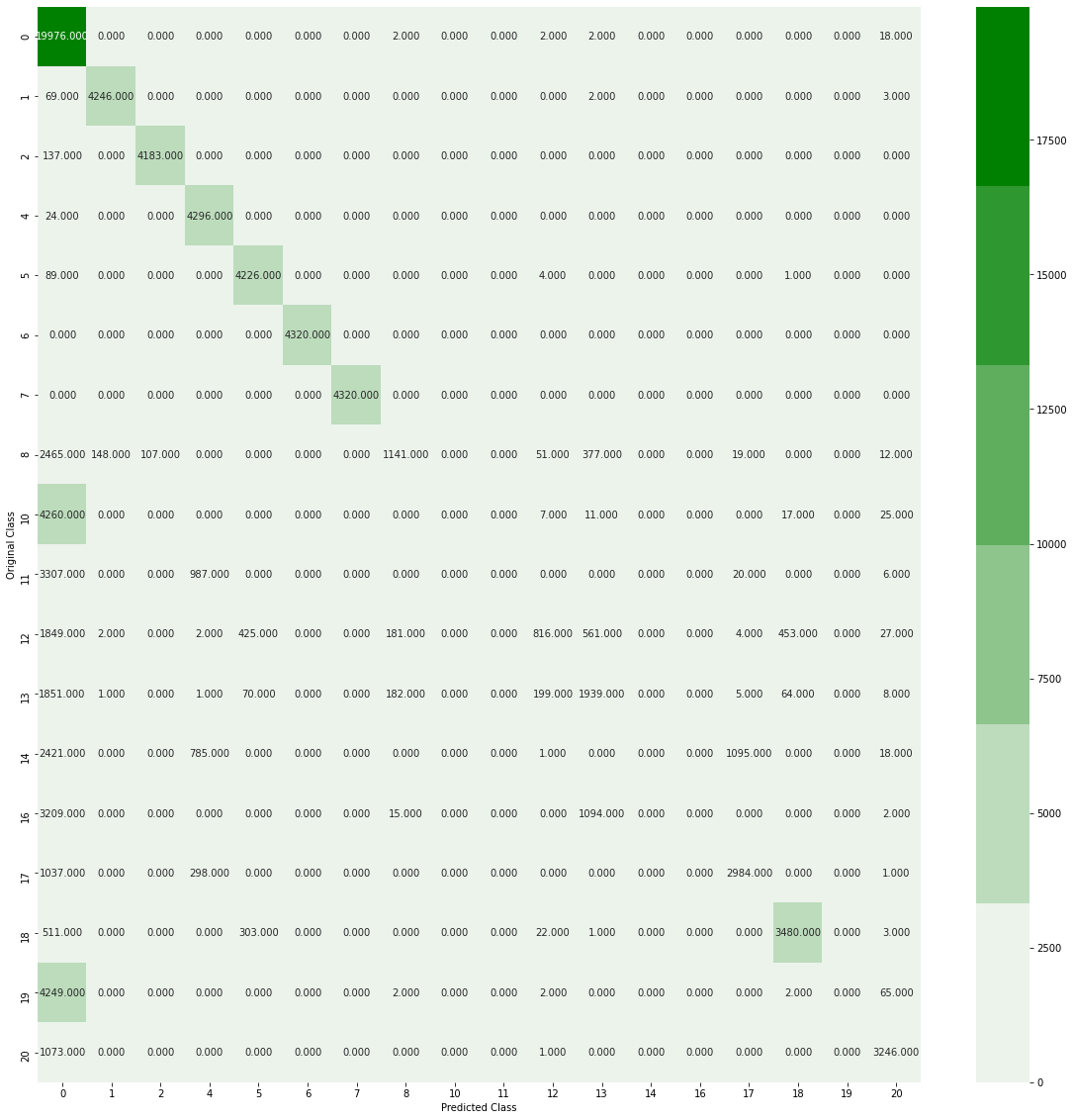}
\caption{Logistic Regression confusion matrix}
\label{logistic_confusion_matrix}
\end{figure}

To supplement Section 4.2, we compare the best model LSTM's confusion matrix (Fig. \ref{confusion_matrix}) against the CNN (Fig. \ref{cnn_confusion_matrix}), Random Forest (Fig. \ref{randomforest_confusion_matrix}), and Logistic Regression (Fig. \ref{logistic_confusion_matrix}).

\newpage
\section*{Appendix C: More on the TE Dataset}
\subsection*{Sampling the TE Dataset}
As the TE dataset itself is too large to train efficiently, we performed sampling on the original dataset to obtain a more manageable subset. A description of this dataset is described below.

The TE dataset is deliberately split into two parts that are similar in terms of size to create a more balanced dataset: fault data and non-fault data. It is furthermore split into test and training sets. So we have 4 files: fault\_training, fault\_testing, non-fault\_training and non-fault\_testing.

Using the training set as the example: there are 21 types for each data point. Type 0 is non-fault and the rest are different types of faults. Faults are introduced after a few hours of normal operation. There are 500 samples of each type of fault. So there will be 500*21 = 10500 data points for each simulation. There are 500 simulations, so the total number of data points is 500*10500 = 5250000. Every simulation is very similar to each other by the nature of TE process.

Our sampling method is borrowed from \cite{sawant_2019}. From fault-training and non-fault\_training, we take the fault 0 (No Fault) data points from the first 40 simulations and take the rest of of fault types data points in the first 25 simulations while ignoring the fault 0 to form our training set. Similarly, we take another 20 simulations for fault 0 and 10 simulations for the rest of fault types to form the validation set. Then, from fault-testing and non-fault\_testing, we take the fault 0 data points from the first 2 simulations and take the rest of fault types from the first 10 simulations to form our testing set.

Note that in our experiment, the machine learning models perform relatively the same on the sampled subset and the original and much larger dataset.

\subsection*{Access to the Dataset}
Please refer to our Github repository (\url{https://github.com/BerkeleyExpertSystemTechnologiesLab/EVSIvsLSTM}) for the sampled dataset in CSV form, code and models.
\end{document}